\documentclass[aps,pra,reprint,showpacs,superscriptaddress]{revtex4-1}
\usepackage{amsmath,times,bm}
\usepackage[dvipdfmx]{graphicx,color,hyperref}
\newcommand\ket[1]{| #1 \rangle}
\newcommand\bra[1]{\langle #1 |}

\begin{document}
%%%%%%%%%%
% information
%%%%%%%%%%
\title{Noncommutative quantum mechanics and skew scattering in ferromagnetic metals}
\date{\today}
\author{Hiroaki Ishizuka}
\affiliation{Department of Applied Physics, The University of Tokyo, Hongo, Bunkyo-ku, 
Tokyo 113-8656, Japan}
\author{Naoto Nagaosa}
\affiliation{Department of Applied Physics, The University of Tokyo, Hongo, Bunkyo-ku, 
Tokyo 113-8656, Japan}
\affiliation{RIKEN Center for Emergent Matter Science (CEMS), Wako 351-0198, Japan}                      
% 72.15.-v 	Electronic conduction in metals and alloys
% 72.15.Gd 	Galvanomagnetic and other magnetotransport effects (see also 75.47.-m Magnetotransport phenomena; materials for magnetotransport)
% 72.20.Dp 	General theory, scattering mechanisms 
\begin{abstract}
  The anomalous Hall effect in ferromagnetic metals is classified into two based on the mechanism. The first one is the intrinsic Hall effect due to the Berry curvature in momentum space; this is a Hall effect that solely arises from the band structure of solids. On the other hand, another contribution to the Hall effect, so-called extrinsic mechanism, comes from impurity scatterings such as skew scattering and side jump; for the extrinsic mechanism, the spin-orbit interaction of the impurity ions is often required. These two mechanisms are often discussed separately; the intrinsic Hall effect is dominant in the intermediate resistivity region while the latter, i.e., skew scattering, becomes important in the clean limit. In this work, it is shown theoretically that the non-commutative nature of the real-space coordinates in the presence of the Berry curvature causes the skew scattering by the nonmagnetic impurity without relativistic spin-orbit interactions, in sharp contrast to the conventional cases.
\end{abstract}
\maketitle
%%%%%%%%%%
% mainmatter
%%%%%%%%%%
Berry phase connection
\begin{eqnarray}
\bm{a}(\bm{k})={\rm i}\bra{u_{\bm k}}\nabla_{\bm k}\ket{u_{\bm k}}
\end{eqnarray}
of the band structures in solids, which describes how the two neighboring Bloch functions are overlapped in the crystal momentum ($\bm{k}$)-space, plays important roles in a variety of phenomena~\cite{Thouless1982,Berry1984,Xiao2010}. ($\ket{u_{\bm k}}$ is the periodic part of the Bloch function with crystal momentum $\bm k$, and $\nabla_{\bm k}$ is the gradient operator with respect to $\bm k$.) This $\bm{a}(\bm{k})$ plays the role of the vector potential, and leads to the Berry curvature $\bm{b}(\bm{k}) = \nabla_{\bm{k}} \times \bm{a}(\bm{k})$ analogous to the magnetic field. The Berry connection  $\bm{a}(\bm{k})$  has the physical meaning of the intracell coordinate~\cite{Resta1992,KingSmith1993}, i.e., the real-space position of the wavepacket measured from the Wannier coordinate reads
\begin{equation}
  \bm{r} = i \frac{\partial}{\partial \bm{k}} + \bm{a}(\bm{k}) \label{realspace}.
\end{equation}
On the other hand, the Berry curvature  $\bm{b}(\bm{k})$ gives a nonzero commutation relation between the components of the real-space coordinate $\bm{r}$. For example, 
\begin{align}
  [x, y] &= [ i \frac{\partial}{\partial k_x} + a_x(\bm{k}), 
   i \frac{\partial}{\partial k_y} + a_y(\bm{k})]  \\
   &= i \biggl[ \frac{\partial a_y }{\partial k_x} - \frac{\partial a_x}{\partial k_y} \biggr]  = i b_z(\bm{k}) \label{commu}.
\end{align}
Therefore, the wavepackets made of the Bloch functions are described by the non-commutative quantum mechanics. This fact leads to the so-called anomalous velocity and also the intrinsic anomalous Hall effect (AHE) in metallic ferromagnets~\cite{Karplus1954,Kohn1957,Luttinger1958,Jungwirth2002,Onoda2002}. Namely, the transverse anomalous velocity to the external electric field is induced by the Berry curvature  $\bm{b}(\bm{k})$, which is the dual to the Lorentz force due to the magnetic field in real space. This intrinsic mechanism due to the geometric nature of the Bloch wavefunctions is now confirmed in many materials by the comparison between the first-principles calculations and experiments~\cite{Miyasato2002,Yao2004,Wang2006,Wang2007,Nagaosa2010}.

Historically, however, the intrinsic mechanism of the AHE was questioned for a long period, as the impurity scatterings relax the momentum distribution to the steady state under the external electric field. As impurity scatterings are inevitable in solids and they seem to cancel the force acting on the electrons, the anomalous velocity induced by the Berry curvature was expected to vanish; thus, no intrinsic AHE. Therefore, the extrinsic mechanisms due to impurity scattering were established earlier. Historically, Smit was the first to propose the extrinsic mechanism of AHE by the skew scattering~\cite{Smit1955,Smit1958}, where the transition probability for the scattering ${\bm k} \to {\bm k}^{\prime}$ is different from that of ${\bm k}^{\prime} \to {\bm k}$, i.e., the detailed balance condition is broken. Later, another extrinsic mechanism called side jump was proposed~\cite{Berger1970}, where a transverse shift of the electron trajectory occurs at the scatterers.

Usually, the intrinsic and extrinsic mechanisms of AHE are discussed separately, and considered to be dominant in different regimes of the longitudinal resistivity $\rho_{xx}$~\cite{Nagaosa2010}; the intrinsic one is dominant in the region $1 \mu \Omega cm < \rho_{xx} < 1 m \Omega cm$ while the skew scattering is dominant for $\rho_{xx} < 1 \mu \Omega cm$. The side jump mechanism is also effective, but often smaller than these two. Technically, skew scattering appears in the second Born approximation~\cite{LerouxHugon1972}; it appears from the interference of the first order and second order scattering processes. In this mechanism, the spin-orbit interaction (SOI) at the scatterer plays a key role in the asymmetry of the scattering amplitude.

In this paper, we study the scattering by a potential {\it without SOI} for the electronic states with finite Berry curvature in terms of the non-commutative quantum mechanics. The key observation is that the nonzero commutators between the components of the real-space coordinates urge to introduce the new canonical coordinates, which satisfies the usual commutation relations [See Eqs.~\eqref{eq:Xdef} below.]. This results in the asymmetric scattering as we see in Eqs.~\eqref{eq:Vimp} and \eqref{eq:Wasym2}, which leads to the skew scattering.

Remarkably, Smit already discussed that the SOI at the impurity potential is not required for the skew scattering~\cite{Smit1958}. When the host electronic states are influenced by the SOI, the scattering matrix element contains the skewness which is typically expressed by 
\begin{equation}
  <\bm{k}| V | \bm{k}'> = i \lambda \bm{\sigma} \cdot ( \bm{k} \times \bm{k}' ) \label{smit}.
\end{equation}
where $\bm{\sigma}$ is the Pauli matrices vector corresponding to the electron spin, $\lambda$ is the coefficient of scattering, typically in the order of the SOI, and $\bm{k}$ and $\bm{k}'$ are the wavevectors before and after the scattering. This is the usual form of the potential for spin-orbit scattering. Our result in Eqs.~\eqref{eq:Vimp} and \eqref{eq:Wasym2} has a similar form to Eq.~\eqref{smit} when $\bm{b}$ is replaced by $\bm{\sigma}$.

{\it Model} --- In this paper, we consider a three-dimensional space denoted by 
$\bm x=(x,y,z)= (x_1,x_2,x_3)$ and its momentums $\bm p=(p_x,p_y,p_z)=(p_1,p_2,p_3)$ with the following commutation relations:
\begin{subequations}
  \begin{align}
    [x_1,x_2]&={\rm i}b,\\
    [x_i,x_3]&=0,\\
    [x_i,p_j]&={\rm i}\delta_{ij},\\
    [p_i,p_j]&=0,
  \end{align}\label{eq:commutation}%
\end{subequations}%
where $i,j=1,2,3$. Throughout this paper, we put $\hbar =1$. To study the effect of impurity scattering, we here consider a single particle Hamiltonian of spinless fermion with (non-magnetic) impurities:
\begin{subequations}
\begin{eqnarray}
  H  &=& H_0 + H_V,\label{eq:Hamil}\\
  H_0&=&\frac{\bm p^2}{2m},\label{eq:H0}\\
  H_V&=&V\sum_i\delta(\bm x-\bm x_i),\label{eq:HV}
\end{eqnarray}
\end{subequations}
where $H_0$ is the Hamiltonian for the free electrons and $H_V$ is the impurity Hamiltonian; $V$ is the strength of potential induced by a scatterer, $\delta(\bm x)$ is the three-dimensional delta function, and $\bm x_i$ is the position of the impurity. The sum in the second term is over all impurities indexed by $i$. Note that $V$ has the dimension of (energy)$\times$(length)$^3$; when we consider the case of impurity atoms replacing the host atoms forming a lattice, $V$ should be replaced by $va^3$, where $v$ is the potential energy and $a$ the lattice constant (Hereafter, we take the unit $a=1$.). In the discussion below, we treat $H_V$ as a perturbation and assume $V$ is the same for all impurities. However, an extension to a set of impurities with different scattering strength is straightforward.

{\it Skew scattering} --- We first investigate the scattering problem with one impurity at the center, i.e., $\bm x_0=\bm 0$. For simplicity, we set $\bm{b} = (0,0,b)$ to be constant. The eigenstates of single particle Hamiltonians with the commutation relation in Eq.~\eqref{eq:commutation} can be obtained by introducing an alternative set of commutative ``position'' operators, $X_1$ and $X_2$, that gives two sets of canonical coordinates and momenta, $(X_1,p_1)$ and $(X_2,p_2)$~\cite{Nair2001}:
\begin{subequations}
  \begin{align}
    X_1&=x_1+\frac{b}2p_y,\\
    X_2&=x_2-\frac{b}2p_x,\\
    X_3&=x_3.
  \end{align}\label{eq:Xdef}
\end{subequations}%
We, here, use this approach to calculate the scattering amplitude of the Hamiltonian in Eq.~\eqref{eq:Hamil}. Using $X_i$ instead of $x_i$ in Eq.~\eqref{eq:commutation}, we obtain three sets of canonical coordinates and momenta:
\begin{subequations}
  \begin{align}
    [X_i,X_j]&=0,\\
    [X_i,p_j]&={\rm i}\delta_{ij},\\
    [p_i,p_j]&=0.
  \end{align}\label{eq:commutationX}
\end{subequations}
Using $X_i$, the impurity Hamiltonian reads 
\begin{eqnarray}
  \bra{\bm k'}V\delta(\bm x)\ket{\bm k} &=& \bra{\bm k'}\left(\frac{V}{(2\pi)^3}\int d\bm q\;e^{ {\rm i}\bm q\cdot\bm x}\right)\ket{\bm k},\\
  &=&\bra{\bm k'}\left(\frac{V}{(2\pi)^3}\int d\bm q\; e^{{\rm i}\bm q\cdot\bm X} e^{{\rm i}\frac{b}2(\bm p\times\bm q)_3}\right)\ket{\bm k},\nonumber\\\\
  &=&e^{{\rm i}\frac{b}2(\bm k\times\bm k')_3},\label{eq:Vimp}
\end{eqnarray}
where $(\cdots)_3$ is the $i=3$ component of the vector in the round bracket. Here, we used Baker-Campbell-Hausdorff formula to factorize the exponential function. As we see in the following calculations, the outer product in Eq.~\eqref{eq:Vimp} gives rise to the skew scattering similar to Eq.~\eqref{smit}.

We calculate the transition probability $W_{\bm k\to\bm k'}$ using Born approximation. Within the second Born approximation, $W_{\bm k\to\bm k'}$ reads
\begin{eqnarray}
  W_{\bm k\to\bm k'} = 2\pi|F^{(1)}(\bm k',\bm k) +F^{(2)}(\bm k',\bm k)|^2\delta(\varepsilon_{\bm k}-\varepsilon_{\bm k'}),\nonumber\\
\end{eqnarray}
where
\begin{eqnarray}
  F^{(1)}(\bm k',\bm k) &=& \langle \bm k'|V\delta(\bm x)|\bm k\rangle,\\
  &=& \frac{V}\Omega e^{{\rm i}\frac{b}2(\bm k\times\bm k')_3},
\end{eqnarray}
and
\begin{eqnarray}
  F^{(2)}(\bm k',\bm k) &=& \langle \bm k'|V\delta(\bm x) G(\bm 0,\varepsilon_k)V\delta(\bm x)|\bm k\rangle,\\
  &=& -\frac{V^2m}{\Omega}\; \frac{e^{ik\left|\frac{b}2(k_3-k'_3)\right|}}{\left|\frac{b}2(k_3-k'_3)\right|},\label{eq:F2}
\end{eqnarray}
are the first and second Born terms, respectively. Here, $|\bm k\rangle$ is the eigenstate for $\bm p$, $\bm p|\bm k\rangle=\bm k|\bm k\rangle$, $\varepsilon_k=k^2/2m$ is the eigenenergy of $|\bm k\rangle$ ($k=|\bm k|$ is the length of vector $\bm k$), $\Omega$ is the volume of the system, and $G(\bm x,\omega)$ is the Green's function for $H_0$,
\begin{eqnarray}
  G(\bm x,\omega)=\int \frac{d\bm k'}{(2\pi)^3}\;
G(\bm k',\omega)e^{ {\rm i}\bm k'\cdot\bm r},\label{eq:Gx}
\end{eqnarray}
where $G(\bm k',\omega)$ is the Fourier transform of $G(\bm x,\omega)$,
\begin{eqnarray}
  G(\bm k',\omega)=\frac1{\omega-\frac{k'^2}{2m}+i\epsilon}\; \frac{\Lambda^2}{k'^2+\Lambda^2}.\label{eq:Gk}
\end{eqnarray}
In Eq.~\eqref{eq:Gk}, $\Lambda$ is the cutoff introduced to avoid the divergence that appears in the integral for $\bm k'$ in Eq.~\eqref{eq:Gx}; we take the $\Lambda\to\infty$ limit at the end of the calculation of $F^{(2)}=(\bm k',\bm k)$. The result in Eq.~\eqref{eq:F2} is after taking the $\Lambda\to\infty$ limit; it turns out $F^{(2)}(\bm k',\bm k)$ converges to a finite value in the limit.

Using the $F^{(1)}(\bm k',\bm k)$ and $F^{(2)}(\bm k',\bm k)$, we calculate the asymmetric part of the scattering rate $W^\text{(asym)}_{\bm k\to\bm k'}$. We find that the leading order of the asymmetric part is $V^3$; it arises from the products $F^{(1)}(\bm k',\bm k)[F^{(2)}(\bm k',\bm k)]^\ast+[F^{(1)}(\bm k',\bm k)]^\ast F^{(2)}(\bm k',\bm k)$. The leading order of $W^\text{(asym)}_{\bm k\to\bm k'}$ reads
\begin{eqnarray}
  W^\text{(asym)}_{\bm k\to\bm k'} &=& \frac12\left(W_{\bm k\to\bm k'}-W_{\bm k'\to\bm k}\right),\\
  &=& -\frac{(2\pi)^3} \Omega\frac{n_iV^3m}{(2\pi)^2} \frac{4w_{\bm k',\bm k}(b)}{\left|b(k_3-k'_3)\right|} \delta(\varepsilon_{\bm k}-\varepsilon_{\bm k'}),\\
  &\sim& -\frac{(2\pi)^3}\Omega\frac{n_iV^3m}{(2\pi)^2}kb(\bm k\times\bm k')_3 \delta(\varepsilon_{\bm k}-\varepsilon_{\bm k'}).\label{eq:Wasym2}
\end{eqnarray}
Here,
\begin{eqnarray}
  w_{\bm k',\bm k}(b)=\sin\left[(b/2)(\bm k\times\bm k')_3\right] \sin\left[(k/2)\left|b(k_3-k'_3)\right|\right],\nonumber\\
\end{eqnarray}
$n_i=N_i/\Omega$ is the density of impurities, $\Omega$ is the volume, and $N_i$ is the number of impurities. In Eq.~\eqref{eq:Wasym2}, we expanded $w_{\bm k',\bm k}(b)$ by $k$ assuming $k_F^2b\ll1$. Eq.~\eqref{eq:Wasym2} has the same $\bm k$ dependence as that of skew scattering induced by an impurity with spin-orbit interaction [Eq.~\eqref{smit}].

{\it Boltzmann Theory} --- We next investigate how the antisymmetric scattering term contributes to AHE. For this, we here consider a semiclassical Boltzmann theory with an antisymmetric scattering amplitude:
\begin{eqnarray}
  q\bm v_{\bm k}\cdot \bm E f'_0(\varepsilon_{\bm k})&=&- \frac{g_{\bm k}}\tau + 
\frac\Omega{(2\pi)^3}\int dk'{}^3 W^\text{(asym)}_{\bm k'\to \bm k}g_{\bm k'},\label{eq:boltzmann_def}\\
  &=& - \frac{g_{\bm k}}\tau\nonumber\\
  &&+ \int d\theta'\sin\theta' d\phi' \frac{\rho(k)}{4\pi}\tilde{\bm V}(k)\cdot\frac{\bm k \times\bm k'}{k^2} g_{\bm k'},\nonumber\\\label{eq:boltzmann}
\end{eqnarray}
where $q$ is the charge of the particle, $\bm E$ is the external d.c. electric field, $\bm v_{\bm k}=\nabla_k\varepsilon_{\bm k}$ is the velocity of the electron in $\bm k$ state, $f'_0(\varepsilon)=d f_0(\varepsilon)/d\varepsilon$ with $f_0(\varepsilon)$ is the Fermi-Dirac distribution function, and $\rho(\varepsilon_k)=mk/2\pi^2$ is the density of states for $H_0$ at energy $\varepsilon_{\bm k}$. We here assumed the occupation of electrons $f_{\bm k}$ is close to $f_0(\varepsilon_{\bm k})$, i.e.,
\begin{eqnarray}
  f_{\bm k}=f_0(\varepsilon_{\bm k})+g_{\bm k},
\end{eqnarray}
where $g_{\bm k}$ is the small deviation from $f_0(\varepsilon_{\bm k})$; the equation is expanded to the linear order in $g_{\bm k}$. In addition, in Eq.~\eqref{eq:boltzmann_def}, we used the relaxation time approximation for the symmetric part of the scattering rate, 
\begin{eqnarray}
  W_{\bm k\to\bm k'}^\text{(sym)}=\frac12\left(W_{\bm k\to\bm k'}+W_{\bm k'\to\bm k}\right),
\end{eqnarray}
i.e., the scattering term that involves $W_{\bm k\to\bm k'}^\text{(sym)}$ is replaced by $-g_{\bm k}/\tau$, where $\tau$ is the relaxation time.

For the integral in Eq.~\eqref{eq:boltzmann}, we assumed the form
\begin{eqnarray}
  W^\text{(asym)}_{\bm k'\to \bm k} = \tilde{\bm V}(k)\cdot\frac{\bm k \times\bm k'}{k^2},
\end{eqnarray}
with $\tilde{\bm V}(k)=[\tilde V_1(k),\tilde V_2(k),\tilde V_3(k)]$ being the function of $k$; this is a generalization of the antisymmetric scattering term in Eq.~\eqref{eq:Wasym2}. The integral is written using the polar coordinate $\bm k'=(k'\cos\theta'\cos\phi',k'\cos\theta'\sin\phi',k'\sin\theta')$; the radius is fixed to $k'=k$ due to the energy conservation, i.e., the delta function in Eq.~\eqref{eq:Wasym2}.

Equation~\eqref{eq:boltzmann}, is solved using a self-consistent approach. For this, we introduce a new parameter
\begin{eqnarray}
  \bm P(k) = \int d\phi'd\theta'\sin\theta'\bm k' g_{\bm k'}.\label{eq:Pk}
\end{eqnarray}
On the other hand, using Eqs.~\eqref{eq:boltzmann} and \eqref{eq:Pk}, $g_{\bm k}$ 
become
\begin{eqnarray}
  g_{\bm k}= -\tau q\bm v_{\bm k}\cdot \bm E f'_0(\varepsilon_{\bm k}) + \frac{\tau \rho(k) \tilde{\bm V}(k)}{4\pi k^2}\cdot\bm k\times\bm P(k).\label{eq:gk}
\end{eqnarray}
Substituting Eq.~\eqref{eq:gk} into $g_{\bm k}$ in the integrand of Eq.~\eqref{eq:Pk}, the solution for $\bm P(k)$ reads
\begin{eqnarray}
  \bm P(k) &=& -\tau q \frac{2\pi k^2}{m}f'_0(\varepsilon_{\bm k}) \frac{\bm E + \frac\tau2 \rho(k)\bm E\times \tilde{\bm V}(k)}{1+\left\{\frac\tau2 \rho(k)\tilde{\bm V}(k)\right\}^2}.
\end{eqnarray}
Therefore, to the leading order in $\bm E$, Eq.~\eqref{eq:gk} reads
\begin{eqnarray}
  g_{\bm k} &=& -\tau q f'_0(\varepsilon_{\bm k}) \bm v_{\bm k}\cdot\left(\bm E + \frac\tau2 \rho(k) \tilde{\bm V}(k) \times \bm E\right).\label{eq:gk2}
\end{eqnarray}
Hence, the contribution from impurity scattering to the transverse conductivity reads
\begin{eqnarray}
  \sigma_{xy}&=&-\frac{nq^2\tau^2}{2m}\rho(\varepsilon_F)\tilde{V}_3(k_F),\label{eq:sxy}
\end{eqnarray}
where $k_F$ is the Fermi velocity and $\varepsilon_F$ is the Fermi energy. For the $W^\text{(asym)}_{\bm k'\to \bm k}$ in Eq.~\eqref{eq:Wasym2}, $\tilde{\bm V}(k)$ reads
\begin{eqnarray}
  \tilde{\bm V}(k) = -2\pi n_iV^3mk^3b\hat{\bm x}_3,
\end{eqnarray}
where $\hat{\bm x}_3=(0,0,1)$ is the unit vector along the $x_3$ axis. Therefore, the transverse conductivity become
\begin{eqnarray}
  \sigma_{xy}&=&\frac{nq^2\tau^2n_i}{2\pi}V^3mk_F^4b.\label{eq:sxy2}
\end{eqnarray}

When the major source of scattering is the elastic scattering by the impurities, $\tau$ in Eq.~\eqref{eq:sxy} is estimated to be $1/\tau\sim n_iV^2\rho(\varepsilon_F)$. On the other hand, from Eq.~\eqref{eq:Wasym2}, we see that the leading order of $\tilde{V}(k)$ reads $\tilde{V}(k)\sim n_iV^3\rho(\varepsilon_F)bk_F^2$. Therefore, similar to the skew scattering by an impurity with SOI, the Hall conductivity is $\sigma_{xy}\sim \rho(\varepsilon_F) V k_F^2b\sigma_{xx}$ with $\sigma_{xx}=nq^2\tau/m$ being the longitudinal conductivity. Hence, the Hall angle for the AHE due to skew scattering is estimated as $\sigma_{xy}/\sigma_{xx}\sim V\rho(\varepsilon_F)k_F^2b$. This result indicates a relation between the longitudinal ($\rho_{xx}$) and transverse ($\rho_{yx}$) resistivities $\rho_{yx}\propto\rho_{xx}$ with the fixed strength of the impurity potential $V$.

In addition to the skew scattering we discussed here, an electronic band with a finite net Berry curvature shows intrinsic AHE~\cite{Karplus1954,Kohn1957,Luttinger1958}; the intrinsic Hall conductivity is proportional to the number of carriers and Berry curvature, $\sigma^\text{(int)}_{xy}\sim nq^2b$. A key difference is that $\sigma^\text{(int)}_{xy}$ is insensitive to the longitudinal conductivity. Therefore, it is expected that, the skew scattering becomes the major source of Hall effect when the system is clean while the intrinsic Hall effect dominates when $\sigma_{xx}$ is small. The crossover occurs when $\sigma_{xy}^\text{(sk)}/\sigma_{xy}^\text{(int)}\sim k_F^3\tau V=1$; this indicates that the crossover of AHE from intrinsic to skew scattering occurs at $\sigma_{xx}\sim q^2/mV$. Therefore, the crossover occurs at a smaller value when $V$ is stronger.

To summarize, in this work, we studied the anomalous Hall effect from the viewpoint of non-commutative quantum mechanics. In presence of the Berry curvature $\bm b(\bm k)$, we find that a non-magnetic impurity without spin-orbit interaction also contributes to the skew scattering. Using a Boltzmann theory, we present the explicit form of the anomalous Hall conductivity induced by this mechanism. Analogous to the case of the skew scattering by an impurity with spin-orbit interaction, the skew scattering in the current mechanism also results in a Hall conductivity that is linearly proportional to the longitudinal conductivity.

In real materials, the Berry curvature in the band structure often arises as a consequence of spin-orbit interaction. Our results here shows that, in the materials where the bands are strongly modified by the spin-orbit interaction, a non-magnetic impurity with negligible spin-orbit interaction can be a source of skew scattering, which then results in an anomalous Hall effect once the electron (pseudo)spin is polarized, e.g., in a ferromagnetic phase.

Our result also implies that the Berry curvature of the electrons are the key quantity that determines the nature of scattering in the weak $V$ limit. When the typical energy scale of the scattering by the impurities ($\sim n_iV$) is much weaker than the typical energy difference of the different bands ($\Delta$), i.e., $n_iV\ll\Delta$, the scattering between different bands are negligible. However, even in this limit, estimating the anomalous Hall effect is nontrivial in a system with strong spin-orbit interaction, as the magnetism induce hybridization between the Kramers pairs due to the spin-orbital entanglement induced by the spin-orbit interactions. Nevertheless, our theory here implies that the asymmetry in the scattering rate for the nontrivial electronic bands with Berry phase can be evaluated solely by the Berry curvature, without any further details on the Wannier function and the hybridization thereof induced by the magnetism.

%%%%%%%%%%
% aknowledgements
%%%%%%%%%%
We thank the useful discussion with K. Misaki. This work was supported by CREST, JST (No.JPMJCR16F1), and JSPS KAKENHI (Grant No. JP26103006 and JP16H06717).

%%%%%%%%%%
% bibliography
%%%%%%%%%%
%\bibliographystyle{plain}

\end{document}